# Techniques for Dealing with Uncertainty in Cognitive Radio Networks

Fatima Salahdine[1,2], Naima Kaabouch[1], Hassan El Ghazi[2]
[1]Electrical Engineering Department, University of North Dakota, Grand Forks, USA
[2]STRS Laboratory, National Institute of Posts and Telecommunication, Rabat, Morocco
Email: fatima.salahdine@und.edu, naima.kaabouch@engr.und.edu, elghazi@inpt.ac.ma

*Abstract*—A cognitive radio system has the ability to observe and learn from the environment, adapt to the environmental conditions, and use the radio spectrum more efficiently. However, due to multipath fading, shadowing, or varying channel conditions, uncertainty affects the cognitive cycle processes, measurements, decisions, and actions. In the observing step, measurements (i.e., information) taken by the secondary users (SUs) are uncertain. In the next step, the SUs make decisions based on what has already been observed using their knowledge bases, which may have been impacted by the uncertainty, leading to wrong decisions. In the last step, uncertainty can affect the decision of the cognitive radio system, which sometimes can lead to the wrong action. Thus, the uncertainty propagation influences the cognitive radio performance. Therefore, mitigating the uncertainty in the cognitive cycle is a necessity. This paper provides a deep overview of techniques that handle uncertainty in cognitive radio networks.

I. INTRODUCTION

Wireless networks have grown exponentially over the last decade and the traffic of information has exponentially increased. This has created a high demand for radio spectrum frequency bandwidth. However, most licensed frequency bands are sparsely used or unused by their owners. The U.S. Federal Communications Commission (FCC) and recent studies have shown that with fixed spectrum allocation policy, frequency band utilization ranges from 15% to 85% [1], which means there are holes in the spectrum. These holes, called white space, are the non-used spectrum by their owner, called primary users (PUs) [2].

In addition to the inefficient utilization of the radio spectrum, the spectrum is a scarce resource. A logical way to overcome the spectrum scarcity is to use it dynamically by sharing the spectrum with other unlicensed users (SUs) without interfering with the transmission of the PUs. This allows SUs to sense unused channels and use them for transmission [3]. The opportunistic spectrum access (OSA) has been proposed as a solution for the spectrum allocation problems. The OSA policy allows the spectrum to be shared with all users in contrast to the fixed spectrum access (FSA) policy, in which the spectrum is divided into numerous bandwidths assigned to one or more dedicated users. Under the FSA policy, PUs have access to some specific spectrum bands to transmit their data while others are forbidden [4]. In order to advance the use of OSA, several solutions have been proposed, including cognitive radio, which is an enabling paradigm for opportunistic spectrum access.

Cognitive radio is an innovative approach to wireless networking in which the radio device is aware of its environment and has the ability to establish and adjust its parameters autonomously. It has the ability to observe and learn from its environment, adapt to the environmental conditions, and make decisions to use the radio spectrum more efficiently [5]. Indeed, a cognitive radio can perform the following processes: (1) sensing, which is the comprehension and awareness of the environment; (2) deciding, which is the analysis of results and reliable decision-making based on what is sensed from the environment; and (3) acting intelligently by adapting, changing, and adjusting radio parameters to enhance the performance and overcome the spectrum scarcity issue.

Cognitive radio cycle has three main phases, observation, decision-making, and taking a decision [6]. The first stage is critical since it is the stage where the measurements are taken and the spectrums ensign is performed. Multipath fading, shadowing from obstacles and varying channel conditions are the resources of uncertainty and randomness, which affects all the cognitive cycle processes [7]. When the SUs observe the spectrum and take uncertainty measurements, this uncertainty will be spread to the next stages and this can lead to wrong decisions based on uncertain measurements. SUs make decisions based on what has already been observed using their knowledge bases which may have impacted by the uncertainty. Wrong actions will be then taken. Thus, the uncertainty spreads in all the cognitive cycle stages from the spectrum sensing to the taken action, thus the cognitive radio performance degrades.

Therefore, there is a great need to address these uncertainty problems in the cognitive cycle by sensing the spectrum correctly, making the correct decision, and taking the right action. Existing spectrum sensing techniques, such as energy detection [8] matched filter detection [9], do not consider the uncertainty when measurements are missing or uncertain due to a number of parameters, namely, noise, channel condition changes, fading, shadowing, or interferences. In addition, unknown channel impulse response (CIH) is also an uncertainty resource. CIH represents the channel behavior and its exact value can only be estimated. In order to estimate the fading level in the channel, Doppler and delay spread are considered and can replace the CIH [10].

In order to handle the uncertainty in the cognitive cycle, a model that considers uncertainty in all stages of the cognition cycle should be developed, in which the handling uncertainty solution is applied to provide reliable decisions, leading to intelligent actions by the cognitive radio systems. There is a need for a model that can consider the uncertainty in all cognitive cycle phases to ensure high performance and significant reliability. This paper is a deep overview of the techniques that can mitigate uncertainty in cognitive radio. These techniques are classified into four main categories: probabilistic, fuzzy set theory, possibility theory, and evidence theory methods. The remainder of this paper is organized as follows. Section II represents the uncertainty classification. Section III reviews the handling uncertainty techniques and their application in cognitive radio. Section IV compares the reviewed techniques. Finally, a conclusion is given at the end.

## II. Uncertainty classifications

According to its origin, uncertainty is classified into two main classes [17], aleatoric and epistemic, as illustrated in Fig. 1.

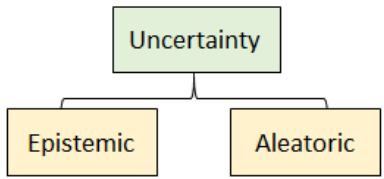

Fig. 1. Uncertainty categories.

In general review, aleatoric uncertainty is a statistical uncertainty that reflects the inherent randomness in nature. It represents unknowns that differ each time the same experiment is done. It cannot be eliminated or predicted by collecting more information or knowledge. The studied system can eventually behave differently depending on this uncertainty. In simple terms, it is simply random [11]. Epistemic uncertainty is a systematic uncertainty that is due to a lack of knowledge and subsequent ability to model and measure the studied system. When data are available, epistemic uncertainty can be presented using probabilities and it can be decreased by collecting more information about the studied system [12]. Both categories exist in real applications. Aleatoric uncertainty arises from stochastic behavior and epistemic uncertainty arises from parameter estimation. The uncertainty type should be first identified in order to mitigate its spread in a specific system. In [13], the two classes were combined in one as a hybrid framework when both are propagated in a dynamic system.

In the context of cognitive radio, we are handling epistemic uncertainty while spectrum sensing. In order to handle the uncertainty and data deficiency and avoid imprecise decisions, several qualification methods have been proposed under the epistemic type [14-17]. These methods are classified into four categories: probabilistic, fuzzy set, possibility, and evidence based theories. Fig. 2 illustrates the classifications of the epistemic uncertainty mitigation techniques.

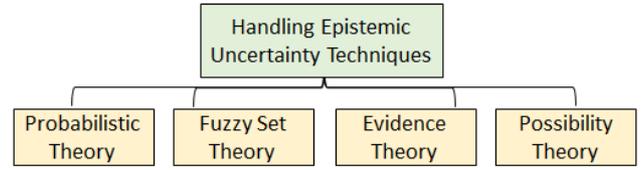

Fig. 2. Representation of methods to handle uncertainty.

Probability theory is the main tool used to estimate all measures of uncertainty. It is a mathematical approach aiming to analyze random phenomena based on random variables, stochastic processes, and events [14]. Fuzzy set theory is an alternative way to handle uncertain and imprecise information to make reliable decisions [15]. Evidence theory is an alternative approach to probabilistic approach for modeling the epistemic uncertainty [16]. Possibility theory is a method to mitigate with uncertainty and incomplete or imprecise data in multisource information [17].

## III. Handling uncertainty techniques

### A. Probabilistic theory based techniques

Probabilistic methods can handle both epistemic through experiments and subjective aleatory uncertainty. Under this category, the degree of belief replaces the knowledge about a system state. The degree of belief is attached to all possible events for the studied system and it is expressed using probabilities since knowledge provides a degree of belief and not certain information. Probabilities relate statements to a state of knowledge. They are expressed as P(*A*/*B*), changing with new evidence *C* to be expressed as P(*A*/*B*, *C*) [14]. Using probabilistic theory allows the studied system to choose the best action. Graphical models are examples of techniques classified under this category. Bayesian network [18] and Markov network [19] are examples of graphical models.

*1) Bayesian Network*

Bayesian network is used to present knowledge about an uncertain domain and model how intervening variables influence one another. It allows a system to handle uncertain contexts and express conditional probabilities of events where data is missing [20]. It is a graphical representation of probabilistic relationships between a set of random variables and their conditional dependencies related via a directed acyclic graph, which make it reflects the conditional relations between variables based on directed links between them. These links indicate direct influence from one variable to another, which is the direct dependence between them. The lack of connection identified the conditional independence between variables.

Bayesian models are based on joint probabilities and conditional probabilities to operate and provide probability of how an event is true. The chain rule is used to write any joint probability distribution as an incremental product of conditional distributions.

$$P(x_1, x_2, \ldots, x_n) = \prod_i P(x_i / x_1, x_2, \ldots, x_{i-1}) \quad (1)$$

where $x_i$ is an event, $i=1\ldots n$, $P(x_i)$ is the prior probability of $x_i$, and $P(x_i/x_j)$ is the conditional probability of $x_i$ given $x_j$. The joint probability distribution $P(X=x)$ can be expressed as a function of conditional probabilities associated with each node $x_i$ under the conditional independence hypothesis.

$$P(X=x) = \prod P(x_i/P_a(x_i)) \quad (2)$$

where $P_a(x_i)$ is the probability of the parent $x_i$ of the child $x_j$ if $x_j$ depends on $x_i$. Bayes' theorem expresses the relations between events using conditional probabilities and it has the form

$$P(x_i/x_j) = P(x_j/x_i) P(x_i) / P(x_j) \quad (3)$$

This theorem computes probabilities when there is not direct information about an event. It allows the representation of causal dependencies between various contextual events and the obtainment of probability distributions.

The Bayesian models have been used in cognitive radio networks to overcome the uncertainty issues in spectrum sensing. The SUs need to sense the band to find the available channel for transmission. The spectrum sensing problem is reformulated as follows

$$y(k) = \begin{cases} n(k) & , H0 \\ \sqrt{E_s}\, e^{j\varphi(k)} + n(k), & H1 \end{cases} \quad (4)$$

where $y(k)$ is the SU received signal, $E_s$ is the PU signal energy, $\varphi(k)=0, \pi$, and $n(k)$ is an Additive White Gaussian Noise (AWGN) signal with zero mean and variance $N_o/2$. Based on Bayesian criterion, decision tests can be written as

$$\left[\frac{P(y/H_1)}{P(y/H_0)}\right] \gtreqless \left[\frac{P(H_0)(C_{pq}-C_{pp})}{P(H_1)(C_{qp}-C_{qq})}\right] \quad (5)$$

where $P(y/H_1)$ and $P(y/H_0)$ are the probability density functions (PDF) of $H_1$ and $H_0$ and $P(H_1)$ and $P(H_0)$ are the prior probability of $H_1$ and $H_0$. $P(H_1)$ and $P(H_0)$ are assumed to be known. The PDF of the SU received signal $y(n)$ over an $N$ symbol duration under the hypotheses $H_0$ and $H_1$, respectively, can be written as follows [21].

$$P(y/H_0) = \prod_{k=0}^{N-1} \frac{e^{-y^2(k)/N_0}}{\sqrt{\pi N_0}} \quad (6)$$

$$P(y/H_1) = \prod_{k=0}^{N-1} \frac{e^{\frac{\left(y(k)-\sqrt{E_s}\,e^{j\varphi(k)}\right)^2}{N_0}}}{\sqrt{\pi N_0}} P(\varphi(k)) \quad (7)$$

Then, from equations (6) and (7), the likelihood ratio test (LRT) of $H_1$ and $H_0$ can be defined as

$$LRT(y) = \frac{P(y/H_1)}{P(y/H_0)} = \prod_{k=0}^{N-1} e^{-E_s/N_0} \cosh(x(k)) \quad (8)$$

where $x(k) = (2\sqrt{E_s})/N_0\, y(k)$. From (7) and (8), the test decision can be deduced by comparing LRT to the threshold δ. The Bayesian detector is based on the Bayesian decision rule, which aims to minimize the expected posterior cost (EPC) [21]. The EPC can be formulated as a function of the cost $C_{ij}$ associated to the decision $H_i$ if the state is $H_j$ for i, j=0, 1

$$EPC = \sum_{i=0}^{1}\sum_{j=0}^{1} C_{ij} P(H_j) P(H_i/H_j) \quad (9)$$

From (10), the optimal Bayesian detector can be derived

$$\sum_{k=0}^{N-1} \ln\left(\cosh(x(k))\right) > N\gamma + \ln(\delta) \quad (10)$$

The threshold is determined using as

$$\delta = \frac{P(H_0)(C_{10}-C_{00})}{(PH_1)(C_{01}-C_{11})} \quad (11)$$

The suboptimal Bayesian detector is obtained using approximations for the low and high SNR. The Bayesian rule is also used to derive expressions for $P_d$ and $P_f$ [22].

Consider a cognitive radio network with $N$ SUs and $M$ licensed channels; the network is modeled by a graph in which each node represents an SU and each edge represents the communication link between the two corresponding nodes (SUs). Bayesian model is used for each SU for decision making and channel selection [23]. In [24], spectrum sensing is reviewed under noise uncertainty caused by the channel conditions, which are realistic time-varying multipath fading channels. The authors propose a spectrum sensing scheme, based on energy detection techniques. The proposed scheme deals with uncertainty by estimating both the PU state and time-variant multipath gains. In [25], the authors proposed a new approach based on BN and game theory, also called Bayesian games, which allow the analysis of the influence of incomplete data and uncertain information in cognitive radio networks. They address the uncertainty about several "players" decision (SUs) in a cooperative system. In [26], a generalized likelihood ratio test (GLRT) is used to address the incomplete knowledge in sensing measurements by considering some parameters known and others unknown and studying the influence and the interactions between them. In [27], in order to learn about the environment conditions, authors proposed a Bayesian network that represents the relationships between variables indicating how the system is performing and identifying which variables affecting bit error rate (BER). Those variables are bit energy to noise spectral density ratio, carrier to interference ratio, modulation scheme, Doppler phase shift, and BER.

*2) Markov models*

Markov models are undirected graphical models that can be categorized in four divisions used in different situations, according to the system observed (autonomous or controlled) and the system state. These models are Markov chain, Hidden Markov model, Markov decision process, and Partially Observable Markov decision process. The Hidden Markov model is a statistical Markov model [19] in which the modeled system is assumed to be a Markovian process with unobserved states. It starts with a Markov chain and adds a noisy observation about the state at each time.

The Markov approach is used in opportunistic spectrum access to model the interactions between PU and SU as continuous-time Markov chains. Hidden Markov models are used to identify the signal features processing in which spectrum activities are Markovian. In [28], the authors proposed a Markov model to obtain the channel holding time, which is the period that the SU uses the free band without PU signal interruption. In [29], a new scheme is proposed to reduce the spectrum consumption: channels are grouped in clusters, only one channel must to be sensed, and the other channels are estimated using the historical states and their correlation with the sensed channel in the same cluster. The Markov model is used to model the influence of historical states on the current state in which the combination of channel correlation and the Markov model reduces the number of channels which must be sensed to improve sensing performance [19].

*B. Fuzzy set theory*

Fuzzy set theory is an alternative approach for handling uncertain and imprecise data to make decisions and allows the expression of real situations with a mathematical model [15]. The fuzzy set theory aims to provide a mathematical model to reasoning under uncertainty in detection and decision making. It describes to what degree a decision is certain and reliable using uncertain and imprecise knowledge based on subjective estimation and expert opinions and experiences. Under fuzzy logic, statement truth is not always clear, but depends on a degree of membership to a set or a class; statement truth takes values between 0 and 1, in contrast to classical logic with two possible values: true (1) or false (0). The degree of membership to this set (i.e., the closeness to 0 or 1) indicates the degree of truth of the statement. Fuzzy logic indicates to what degree $X$ is in various sets. The degree of membership in the interval [0, 1] is associated with each element in the fuzzy set [30].

A fuzzy inference system is implemented following a four-stage procedure: fuzzification, fuzzy inference engine, fuzzy rule base, and defuzzification. The system's inputs are the measurements that represent uncertainty; inputs are forwarded to determine the membership function in the fuzzification stage in which the fuzzy rule IF-THEN is then applied to determine the new outputs set. The fuzzified measurements are used in the next stage, and all rules are aggregating using a fuzzy logic operation (e.g., OR, AND, Union, Intersection). In the final stage, the fuzzy set is converted to a defuzzified set which corresponds to the chance value. Fuzzy probability theory is an extension to handle mixed probabilistic/non-probabilistic uncertainty [31].

Fuzzy logic is applied to cognitive radio networks in order to make decisions under uncertain environments. SUs make decisions based on information available from the spectrum sensing stage, which includes measurements, incomplete data, and their own knowledge. SUs decide to whether transmit their data depending on what is sensed and then select which channel frequency to use without creating interference to the PUs. The fuzzy logic-based approach can be used to make decisions under incomplete and doubtful data; it can deal with uncertainty in data by transforming imprecise data into precise data based on fuzzy set inference [32]. SUs need specific information about the channel state and this information depends on the spectrum sensing method used in the sensing stage. A fuzzy-based spectrum handoff is proposed in [33] in order to avoid interference to the PUs. The fuzzy inference system estimates the distance between SU and PU and the required SU power for avoiding interference. The results show that fuzzy logic outperforms classical spectrum sensing. In [34], a fuzzy power control scheme is proposed in which a transmit power control system is designed using a fuzzy logic system to permit dynamic control of the transmitted signal power depending on the interference level experienced by PUs. In [35], fuzzy conditional entropy maximization is used to design energy detection for cooperative spectrum sensing with the aim to minimize the uncertainty in the threshold selection.

In [36], the authors proposed cooperative spectrum sensing based on the fuzzy integral theory in the context of cognitive radio networks. The proposed scheme seeks to handle the uncertainty in the information provided by the local SUs.

*C. Evidence theory*

Evidence theory, also called the theory of belief functions or Dempster–Shafer theory (DST), was developed as an alternative technique to probability theory [16]. It is a mathematical theory for reasoning and modeling the epistemic uncertainty that reasons with belief and plausibility. Belief provides all evidence available for a specific hypothesis; plausibility offers all evidence which is consistent with this hypothesis. The two concepts give an interval of probabilities, including the true probability, with some certainty. In other words, it combines distinct evidence from several sources to compute the probability of the hypothesis. This evidence is represented by a mathematical function called the belief function. This function considers all evidence available for the hypothesis; a hypothesis and its negation do not have any causal relationship between them, which proves that an ignorance of belief does not involve disbelief, but reflects a state of uncertainty. In order to represent uncertainty in a hypothesis, uncertainty is replaced with belief or disbelief as evidence, which is the concept behind DST. This method suffers from high mathematical complexity and needs all possible states to calculate the exact probability while representing good certainty with additional information about the degree of belief. The DST rule for combination can be defined as the procedure for combining distinct states of evidence. The DST rule for combination consists of a space of mass $W$ and the belief mass as a function denoted $m$.

$$m: 2_X \rightarrow [0, 1] \qquad (12)$$

where $X$ is the set that includes all possible states and $2X$ is the set of all the subsets of $X$. DST is able to handle uncertainty, imprecision, ignorance, and lack of data because it is based on the estimation of imprecision and conflict from several sources. In [19], evidence theory was used to deal with the uncertainty in a mixed aleatory-epistemic model.

DST is applied in the cognitive radio context to handle uncertainty and an imperfect knowledge in cognition cycles. In [37], the authors proposed a distributed spectrum sensing model to improve detection and ensure a reliable and credible spectrum sensing decision. The proposed scheme used the DST to combine the local decisions from several SUs to make the final decision. Each local decision mi is associated with a credibility parameter $m_i(c)$ which quantifies the channel condition and sends the information to the access point to make the final decision about the state of the observed band.

It was shown that the proposed scheme performed well in terms of the probability of detection $P_d$ and the probability of false alarm $P_f$ based on the credibility for hypotheses $H_0$ and $H_1$ [37]. In [38], a cooperative model was proposed to enhance the reliability of spectrum detection. Energy detection with a double threshold is used for spectrum sensing and the DST is used to make decisions.

## D. Possibility theory

Possibility theory is a technique to handle certain types of uncertainty, based on fuzzy set concepts. It is an alternative to probability theory; and an extension to fuzzy logic theory. It permits handling of uncertainty in multisource information [17]. It is characterized by the possibility *Poss* measures, which assign numbers to each subset *W* based on fuzzy logic [39]. Possibility theory rules different from probability theory rules; possibility distribution is defined as

$$Poss: _{W \rightarrow [0, 1]}, \max (A \in W, Poss (A) = 1) \quad (13)$$

Conditional independence in possibility theory allows modeling the dependence between uncertain variables as in probability theory, and can be defined as

$$Poss (A, B/C) = Poss (A/C) \otimes Poss (B/C) \quad (14)$$

In order to model the available information, possibility and necessity measures are considered based on possibility distribution. Possibility theory is sued to handle only epistemic uncertainty, and it is a special case or a subset of evidence theory. Possibility logic was applied in many filed including signal processing to handle noise uncertainty in signal propagation. Noise nature cannot be assumed, estimating the noise level in each sample based on some signal characteristics was treated by applying possibility theory which considered the lack of knowledge about noise; noise uncertainty is due to random perturbation propagated through the transmission channel [40]. Theory of possibility was applied on cognitive radio network as a special case of theory of evidence already reviewed, especially for optimization problems.

## IV. METHODS COMPARISON

Several techniques to handle the uncertainty propagation were reviewed. Selecting a solution depends on several criteria, requirements, conditions; and especially the trade-off between precision and complexity. Through analyzing the different techniques, we can conclude that methods that offer good precision represent high complexity. All techniques had their weaknesses and strengths. The choice of which technique to use depends on the problem to be solved.

Probabilistic theory is a powerful tool of handling uncertainty if precise uncertainty is required. The Bayesian network as a probabilistic method is more efficient and it is the most used to present, reason, and model uncertainty. However, it needs prior probability distributions because in practice measurements and parameters are not always Gaussian. Fuzzy theory is easy to implement, it is more suitable for problems that didn't require precise knowledge of uncertainty, but it handled a degree of truth and not the uncertainty. For evidence theory, it provided additional information about the degree to which information is available. DST is more general than possibility and probability theory. There are other approaches that combine various techniques as a tool to mitigate the hybrid uncertainty. Table I summarized and compared the listed techniques.

TABLE I TECHNIQUES COMPARISON

| Probability theory | Fuzzy logic theory | Evidence theory | Possibility theory |
|---|---|---|---|
| -Deal with randomness and subjective uncertainty -Handle epistemic and aleatory uncertainty through experiments | -Deal with imprecise vague information -Describe o what degree a decision is certain and reliable using uncertain and imprecise knowledge | -Deal with epistemic ncertainty - Used where there is some degree of ignorance (incomplete model) | Deal with incomplete or imprecise data in multisource information |
| -Powerful for handling uncertainty in case precise uncertainty is required -Based on mathematical probability to predict an event | -Precision and stability not guaranteed -Performance measured a Posterior -Based on user experience and interpretation | More Flexible | Improves the precision of evidence |
| -Applied to all systems -Complex data handling -Inexac / incorrect | -Applied to systems that are difficult to model - Easy to implement and interpret | High complexity | -Computationally simple -Complexity close to that of classical logic |
| One valued approaches (Truth is one-valued) | Set valued approaches (Truth is many-valued) | Two valued approaches (Truth is 2-valued) | Two valued approaches (Truth is 2-valued) |
| Degree of belief | Degree of membership to set | Belief and Plausibility | Possibility measures |

Understanding uncertainty in the context of the cognitive radio network can be considered the most important task in modeling this uncertainty and mitigating it, identifying its type, its source and its influence on the cognitive radio system. As reviewed, Bayesian models and fuzzy logic are the most used in spectrum sensing for reasoning under uncertainty and making decision.

## V. CONCLUSION

In order to improve access to the limited radio spectrum, DSA is proposed as a solution to overcome the spectrum scarcity problem. Spectrum sensing is the main task of cognitive radio permitting opportunistic access to the spectrum based on decisions made by the SUs. However, due to the randomness features of communication channels, uncertainty impacts all the processes of the cognitive cycle, including the sensing results at SUs receivers. Making decisions under uncertainty is a challenge that cognitive radio systems must face. This paper examined several models and methods to deal with uncertainty. Among these are a probabilistic theory, fuzzy set theory, possibility theory, and evidence theory. Each of these has advantages and disadvantages. The overarching objective of all these methods is to overcome the error made by the detector due to fading, shadowing, and uncertainty noises, to minimize the sensing error, and to enhance the fidelity of sensing outcome and the sensing decision.